\documentclass[twocolumn,prl,showpacs]{revtex4}%
\usepackage{graphicx}%
\usepackage{amsmath}%
\usepackage{amsfonts}%
\usepackage{amssymb}
\usepackage{bm}

\def\e{{\epsilon}}
\def\k{{ {\bm k} }}

\def\q{{ {\bm q} }}
\def\Q{{ {\bm Q} }}

\def\w{{\omega}}
\def\a{{\alpha}}

\begin{document}
\title{Orbital Fluctuation Mediated Superconductivity in Iron Pnictides:\\
Analysis of Five Orbital Hubbard-Holstein Model
}
\author{Hiroshi \textsc{Kontani}$^{1}$, and Seiichiro \textsc{Onari}$^{2}$}

\date{\today }

\begin{abstract}
In iron pnictides, we find that the moderate electron-phonon interaction 
due to the Fe-ion oscillation can induce the critical $d$-orbital fluctuations, 
without being prohibited by the Coulomb interaction.
These fluctuations give rise to the strong pairing interaction for the
$s$-wave superconducting (SC) state without 
sign reversal ($s_{++}$-wave state), which is consistent with 
experimentally observed robustness of superconductivity against impurities.
When the magnetic fluctuations due to Coulomb interaction are also strong,
the SC state shows a smooth crossover from the $s$-wave 
state with sign reversal ($s_\pm$-wave state) to the $s_{++}$-wave state 
as impurity concentration increases.
\end{abstract}

\address{
$^1$ Department of Physics, Nagoya University and JST, TRIP, 
Furo-cho, Nagoya 464-8602, Japan. 
\\
$^2$ Department of Applied Physics, Nagoya University and JST, TRIP, 
Furo-cho, Nagoya 464-8602, Japan. 
}
 

\sloppy

\maketitle


The mechanism of high-$T_{\rm c}$ superconductivity 
in iron pnictides has been an important open problem.
By considering the Coulomb interaction at Fe-ions,
antiferromagnetic (AFM) fluctuation mediated
fully-gapped sign-reversing $s$-wave state ($s_\pm$-wave state) 
is expected theoretically \cite{Kuroki,Mazin}.
Regardless of the beauty of the mechanism,
there are several serious discrepancies for the $s_\pm$-wave state.
For example, although $s_{\pm}$-wave state is expected to be very fragile
against impurities due to the interband scattering \cite{Onari-impurity},
the superconducting (SC) state is remarkably robust against impurities 
\cite{Sato-imp} and $\a$-particle irradiation \cite{irradiation}.
Moreover, clear ``resonance-like'' peak structure observed by 
neutron scattering measurements \cite{christianson} is 
reproduced by considering the strong correlation effect via quasiparticle 
damping, without the necessity of sign reversal in the SC gap
 \cite{Onari-resonance}.
These facts indicate that 
a conventional $s$-wave state without sign reversal ($s_{++}$-wave state)
is also a possible candidate for iron pnictides.

Then, a natural question is whether the electron-phonon ($e$-ph) 
interaction is important or not.
Although first principle study predicts small 
$e$-ph coupling constant $\lambda\sim0.21$ \cite{lambda-LDA},
several experiments indicate the significance of $e$-ph interaction.
For example, the structural transition temperature $T_{\rm S}$ is
higher than the Neel temperature in underdoped compounds, 
although the structural distortion is small.
Also, prominent softening of shear modulus is observed
towards $T_{\rm S}$ or $T_{\rm c}$ in Ba122 \cite{softening}.
Raman spectroscopy \cite{Raman} also
indicates larger $e$-ph interaction.

Interestingly, there are several ``high-$T_{\rm c}$'' compounds 
with nodal SC gap structure, 
like BaFe$_2$(As$_{1-x}$P$_x$)$_2$
 \cite{AsP} and some 122 systems \cite{Izawa}.
Although nodal $s_\pm$-wave state can appear in the spin-fluctuation
scenario due to the competition between the dominant $\Q=(\pi,0)$
and subdominant fluctuations \cite{Kuroki,Scalapino},
the $T_{\rm c}$ is predicted to be very low.
Thus, it is a crucial challenge 
to explain the rich variety of the gap structure 
in high-$T_{\rm c}$ compounds.

In this letter, we introduce the five-orbital Hubbard-Holstein (HH) model
for iron pnictides, considering the $e$-ph interaction by Fe-ion vibrations.
We reveal that a relatively small $e$-ph interaction ($\lambda\lesssim0.3$) 
induces the large orbital fluctuations,
which can realize the high-$T_{\rm c}$ $s_{++}$-wave SC state.
Moreover, the orbital fluctuations are {\it accelerated} by Coulomb interaction.
In the presence of impurities, the $s_{++}$-wave state dominates the 
$s_\pm$-wave state for wide range of parameters.

\begin{figure}[!htb]
\includegraphics[width=0.9\linewidth]{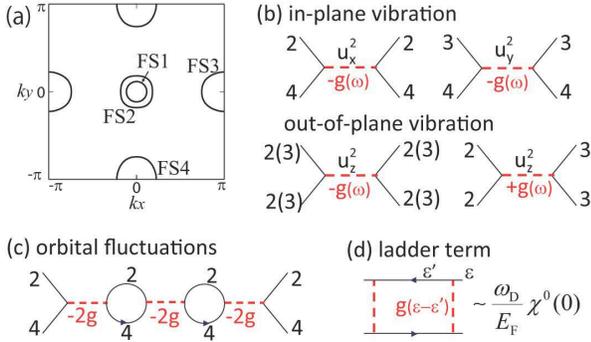}
\caption{
(Color online) 
(a) FSs in the unfolded Brillouin zone.
(b) Phonon-mediated electron-electron interaction.
(c) A bubble-type diagram that induces the critical 
orbital fluctuations between (2,4) orbitals.
(d) A ladder-type diagram that is ignorable when $\w_{\rm D}\ll E_{\rm F}$.
}
\label{fig:V}
\end{figure}

First, we derive the $e$-ph iteration term, considering only
Einstein-type Fe-ion oscillations for simplicity.
Here, we describe the $d$-orbitals in the $XYZ$-coordinate \cite{Kuroki},
which is rotated by $\pi/4$ from the 
$xyz$-coordinate given by the Fe-site square lattice:
We write $Z^2$, $XZ$, $YZ$, $X^2{\mbox{-}}Y^2$, and $XY$ orbitals
as 1, 2, 3, 4, and 5, respectively \cite{Kuroki}.
We calculate the $e$-ph matrix elements
due to the Coulomb potential, by following Ref. \cite{Yada}.
The potential for a $d$-electron at ${\bm r}$ 
(with the origin at the center of Fe-ion)
due to the surrounding As$^{3-}$-ion tetrahedron is
$U^\pm({\bm r}; {\bm u})=3e^2\sum_{s=1}^4|{\bm r}+{\bf u}-{\bm R}_s^\pm|^{-1}$,
where ${\bm u}$ is the displacement vector of the Fe-ion, and
${\bm R}_s^\pm$ is the location of surrounding As-ions;
$\sqrt{3}{\bm R}_s^{+}/R_{\rm {Fe-As}}=(\pm\sqrt{2},0,1)$
and $(0,\pm\sqrt{2},-1)$ for Fe$^{(1)}$, and
$\sqrt{3}{\bm R}_s^{-}/R_{\rm {Fe-As}}=(\pm\sqrt{2},0,-1)$
and $(0,\pm\sqrt{2},1)$ for Fe$^{(2)}$ in the unit cell with two Fe-sites.
Note that $u_{X,Y}$ and $u_Z$ belong to $E_{\rm g}$ and $B_{1g}$ phonons
 \cite{Raman}.
The $\bm u$ linear term of $U^\pm$, which gives the $e$-ph interaction, 
is obtained as
$V^\pm({\bm r}; {\bm u})=\pm A[2XZ\cdot u_X-2YZ\cdot u_Y +(X^2-Y^2)u_Z ]
+O({\bm r}^4)$,
%
%
%
where $A=30e^2/\sqrt{3}R_{\rm {Fe-As}}^4$.
Then, its nonzero matrix elements are given as
\begin{eqnarray}
\langle 2|V|4\rangle=\pm 2a^2Au_X/7,
 \ \ 
\langle 3|V|4\rangle=\pm 2a^2Au_Y/7,
  \nonumber \\
\langle 2|V|2\rangle=\pm 2a^2Au_Z/7,
 \ \ 
\langle 3|V|3\rangle=\mp 2a^2Au_Z/7,
 \label{eqn:Vmat}
\end{eqnarray}
where $a$ is the radius of $d$-orbital.
Here, we consider $\langle i|V|j\rangle$ only for orbitals $i,j=2\sim4$ that 
compose the Fermi surfaces (FSs) in Fig. \ref{fig:V} (a) \cite{Kuroki}.
The obtained $e$-ph interaction does not couple to the charge density
since $\langle i|V|j\rangle$ is trace-less.
Thus, the Thomas-Fermi screening for the coefficient $A$ is absent.
The local phonon Green function is 
$D(\w_l)=2{\bar u}_0^2\w_{\rm D}/(\w_l^2+\w_{\rm D}^2)$, 
which is given by the Fourier transformation of 
$\langle T_\tau u_\mu(\tau)u_\mu(0)\rangle$ ($\mu=X,Y,Z$).
${\bar u}_0= \sqrt{\hbar/2M_{\rm Fe}\w_{\rm D}}$
is the position uncertainty of Fe-ions,
$\w_{\rm D}$ is the phonon frequency, and
$\w_l=2\pi l T$ is the boson Matsubara frequency.
Then, for both Fe$^{(1)}$ and Fe$^{(2)}$, 
the phonon-mediated interaction 
is given by 
%
\begin{eqnarray}
& &V_{24,42}=V_{34,43}=-(2Aa^2/7)^2D(\w_l)\equiv -g(\w_l),
 \nonumber \\
& &V_{22,22}=V_{33,33}=-V_{22,33}=-g(\w_l),
 \label{eqn:Vph}
\end{eqnarray}
as shown in Fig. \ref{fig:V} (b).
Note that $V_{ll',mm'}$ is symmetric with respect to
$l\leftrightarrow l'$, $m\leftrightarrow m'$, and
$(ll')\leftrightarrow (mm')$.
We obtain $g(0)\approx0.4$ eV if we put 
$R_{\rm Fe-As}\approx2.4$ ${\buildrel _{\circ} \over {\mathrm{A}}}$, 
$a\approx0.77$ ${\buildrel _{\circ} \over {\mathrm{A}}}$
(Shannon crystal radius of Fe$^{2+}$), 
and $\w_{\rm D}\approx0.018$ eV.
We have neglected the $e$-ph coupling due to $d$-$p$ hybridization \cite{Yada}
considering the modest $d$-$p$ hybridization in iron pnictides \cite{Singh}.
Thus, we obtain the multiorbital HH model for iron pnictides
by combining eq. (\ref{eqn:Vph}) with the on-site Coulomb interaction;
the intra- (inter-) orbital Coulomb $U$ ($U'$), 
Hund coupling $J$, and pair-hopping $J'$.

Now, we study the rich electronic properties realized 
in the multiorbital HH model \cite{gunnarsson}.
The irreducible susceptibility in the five-orbital model is given by
$\chi^0_{ll',mm'}(q)=-(T/N)\sum_kG_{lm}^0(k+q)G_{m'l'}^0(k)$,
where ${\hat G}^0(k)=[i\e_n+\mu-{\hat H}_\k^0]^{-1}$ is the $d$-electron
Green function in the orbital basis:
$q=(\q,\w_l)$, $k=(\k,\e_n)$, and 
$\e_n=(2n+1)\pi T$ is the fermion Matsubara frequency.
$\mu$ is the chemical potential, and ${\hat H}_\k^0$ is the kinetic term
given in Ref. \cite{Kuroki}.
Then, the susceptibilities for spin and charge sectors 
in the random-phase-approximation (RPA) are given as \cite{Takimoto}
\begin{eqnarray}
{\hat \chi}^{s(c)}(q)=
{\hat \chi}^0(q)[1-{\hat \Gamma}^{s(c)}{\hat \chi}^0(q)]^{-1}.
 \label{eqn:chiSC} 
\end{eqnarray}
For the spin channel,
$\Gamma_{l_1l_2,l_3l_4}^s=U$, $U'$, $J$, and $J'$ for
$l_1=l_2=l_3=l_4$, $l_1=l_3\ne l_2=l_4$, $l_1=l_2\ne l_3=l_4$, and
$l_1=l_4\ne l_2=l_3$, respectively \cite{Kuroki}.
For the charge channel,
${\hat \Gamma}^c=-{\hat C}-2{\hat V}(\w_l)$,
where ${\hat V}(\w_l)$ is given in eq. (\ref{eqn:Vph}),
and $C_{l_1l_2,l_3l_4}=U$, $-U'+2J$, $2U'-J$, and $J'$ for
$l_1=l_2=l_3=l_4$, $l_1=l_3\ne l_2=l_4$ , $l_1=l_2\ne l_3=l_4$, and
$l_1=l_4\ne l_2=l_3$, respectively \cite{Kuroki}.
Figure \ref{fig:V} (c) shows one of bubble diagrams for (2,4)-channel
due to the ``negative exchange coupling $V_{24,42}$''
that leads to a critical enhancement of ${\hat \chi}^c(q)$
 \cite{comment}.
We neglect the ladder diagrams given by ${\hat V}(\w_l)$
in Fig. \ref{fig:V} (d)
since $\w_{\rm D} \ll W_{\rm band}$ \cite{lambda-LDA,Raman}.
We put $\w_{\rm D}=0.02$ eV, $U'/U=0.69$, $J/U=0.16$ and $J=J'$,
and fix the electron number $n=6.1$ (10\% electron doping); the
density of states per spin is $N(0)=0.66$ [eV$^{-1}$].
Numerical results are not sensitive to 
these parameters.
We use $128^2$ $\k$-meshes, and 512 Matsubara frequencies.
Hereafter, the unit of energy is eV.

\begin{figure}[!htb]
\includegraphics[width=0.9\linewidth]{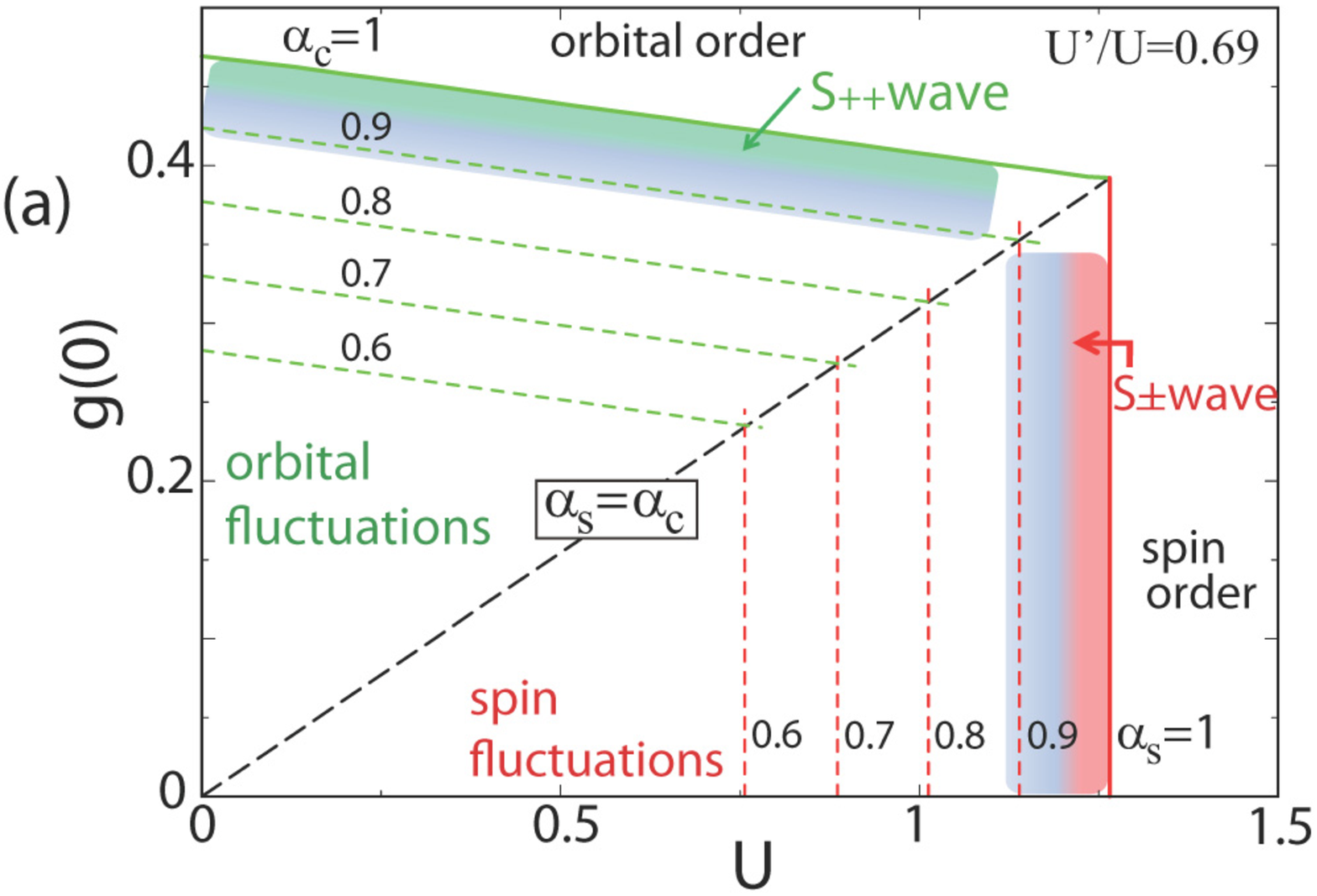}
\includegraphics[width=0.49\linewidth]{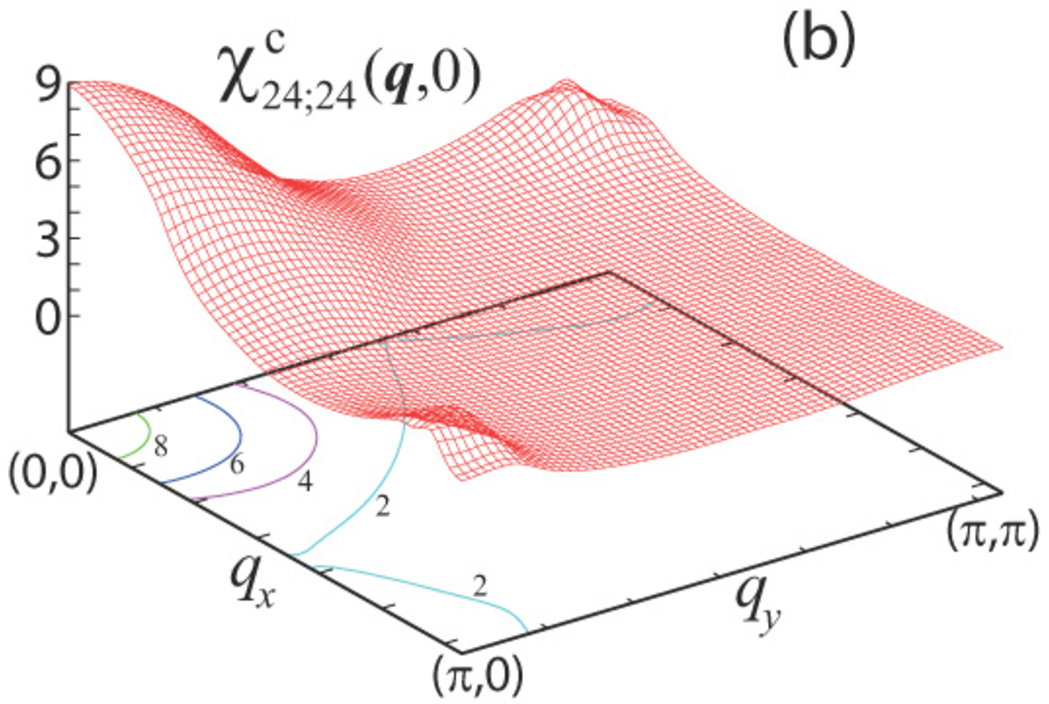}
\includegraphics[width=0.49\linewidth]{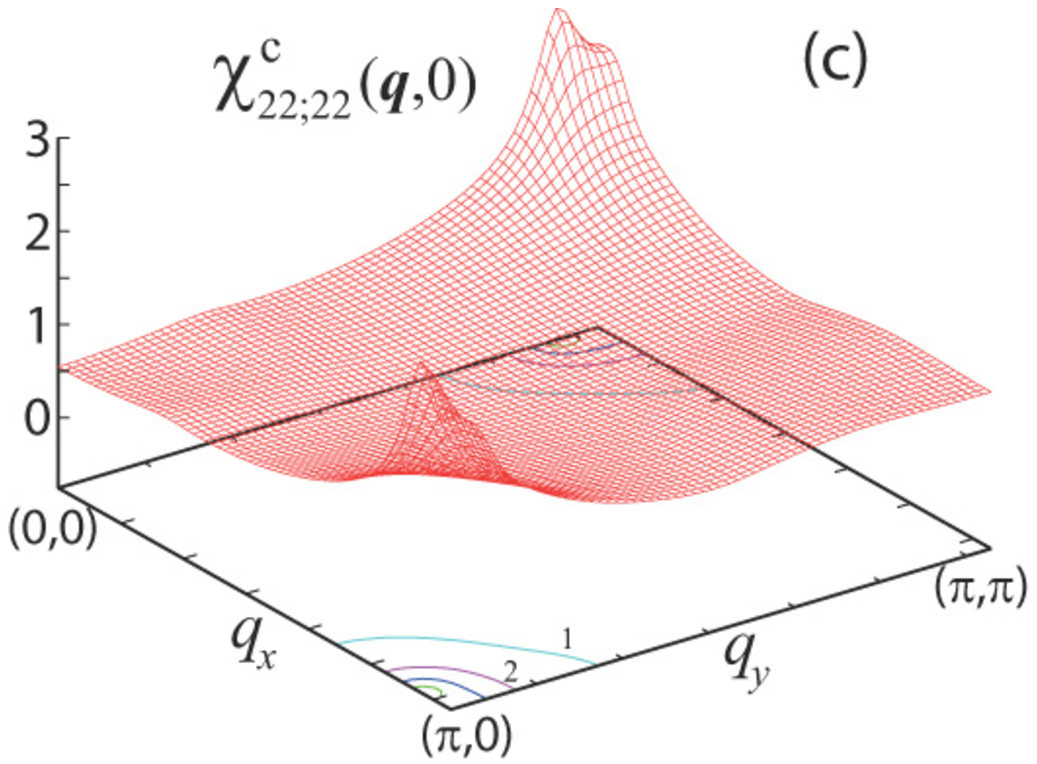}
\caption{
(Color online) 
(a) Obtained $U$-$g(0)$ phase diagram.
(b) Obtained $\chi_{24,42}^c(\q,0)$ and $\chi_{22,22}^c(\q,0)$ for 
$\a_{\rm c}=0.97$.
}
\label{fig:phase}
\end{figure}

Figure \ref{fig:phase} (a) shows the obtained $U$-$g(0)$ phase diagram.
$\a_{\rm s(c)}$ is the spin (charge) Stoner factor, 
given by the maximum eigenvalue of 
${\hat \Gamma}^{s(c)}{\hat \chi}^0(\q,0)$.
Then, the enhancement factor for $\chi^{\rm s(c)}$ is $(1-\a_{\rm s(c)})^{-1}$,
and $\a_{\rm s(c)}=1$ gives the spin (orbital) order boundary.
Due to the nesting of the FSs, the AFM fluctuation with 
$\Q\approx(\pi,0)$ 
develops as $U$ increases, and $s_\pm$-wave state is realized for 
$\a_{\rm s}\lesssim1$ \cite{Kuroki}.
In contrast, we find that the orbital fluctuations develop as 
$g(0)$ increases.
For $U=1$, the critical value $g_{\rm cr}(0)$ for $\a_{\rm c}=1$ is 0.4,
and the critical $e$-ph coupling constant is 
$\lambda_{\rm cr}\equiv g_{\rm cr}(0)N(0)=0.26$ \cite{comment3}.
Since the obtained $\lambda_{\rm cr}$ 
is close to $\lambda$ given by the first principle study 
\cite{lambda-LDA},
strong orbital fluctuations are expected to occur in iron pnictides.
At fixed $U$, $\lambda_{\rm cr}$ decreases as $J/U$ approaces zero.


Figure \ref{fig:phase} (b) and (c) show the obtained
$\chi^c_{ll',mm'}(\q,0)$ for $(ll',mm')=(24,42)$ and $(22,22)$, respectively, 
for $U=1.14$ and $\a_{\rm c}=0.97$ ($g(0)=0.40$):
Both of them are the most divergent channels for electron-doped cases.
The enhancement of $(24,42)$-channel
is induced by the multiple scattering by $V_{24,42}$.
The largest broad peak around $\q=(0,0)$ originates from the 
forward scattering in the electron-pocket (FS3 or 4) 
composed of $2\sim4$ orbitals.
(FS1,2 are composed of only 2 and 3 orbitals.)
These ferro-orbital fluctuations would induce
the softening of shear modulus \cite{softening}, and also 
reinforce the ferro-orbital-ordered state below $T_{\rm S}$ \cite{Shimo-OO}
that had been explained by different theoretical approaches \cite{OO-theory}:
The divergence of $\chi^c_{24,42}$ ($\chi^c_{34,43}$) pushes the 
2,4 (3,4) orbitals away from the Fermi level, and the Fermi surfaces
in the ordered state will be formed only by 3 (2) orbital,
consistently with ref. \cite{Shimo-OO}.
The lower peak around $\Q=(\pi,0)$ comes from the nesting 
between hole- and electron-pockets.
Also, the enhancement of $(22,22)$-channel for $\Q=(\pi,0)$
is induced by the nesting
via multiple scattering by $V_{22,22}$ and $V_{22,33}$.
In contrast, the charge susceptibility
$\sum_{l,m}\chi^c_{ll,mm}(\q,0)$ is finite even if $\a_{\rm c}\rightarrow1$
since $\chi^c_{22,33} \approx -\chi^c_{22,22}$.

Now, we will show that 
large orbital fluctuations, which are not considered
in the first principle study of $T_{\rm c}$ \cite{lambda-LDA},
can induce the $s_{++}$-wave state when $g(0)>0$.
We analyze the following linearized Eliashberg equation 
using the RPA \cite{Kuroki},
by taking both the spin and orbital fluctuations into account
on the same footing:
\begin{eqnarray}
\lambda_{\rm E}\Delta_{ll'}(k)&=&\frac{T}{N}\sum_{k',m_i} W_{lm_1,m_4l'}(k-k')
 \nonumber \\
& &\times G_{m_1m_2}(k')\Delta_{m_2m_3}(k')G_{m_4m_3}(-k'),
 \label{eqn:Eliash}
\end{eqnarray}
where 
${\hat W}(q)=-\frac32{\hat \Gamma}^s{\hat \chi}^s(q){\hat \Gamma}^s
+\frac12{\hat \Gamma}^c{\hat \chi}^c(q){\hat \Gamma}^c 
-\frac12({\hat \Gamma}^s-{\hat \Gamma}^c)$
for singlet states.
The eigenvalue $\lambda_{\rm E}$ increases as $T\rightarrow0$, and
it reaches unity at $T=T_{\rm c}$.
In addition, we take the impurity effect into consideration
since many iron pnictides show relatively large residual resistivity.
Here, we assume the Fe site substitution, where the impurity potential 
$I$ is diagonal in the $d$-orbital basis \cite{Onari-impurity}.
Then, the $T$-matrix in the normal state is given by
${\hat T}(\e_n)= [I^{-1}-N^{-1}\sum_\k{\hat G}(\k,\e_n)]^{-1}$
in the orbital basis \cite{Onari-impurity}.
Then, the normal self-energy is 
${\hat \Sigma}^n(\e_n)= n_{\rm imp}{\hat T}(\e_n)$,
where $n_{\rm imp}$ is the impurity concentration.
Also, the linearized anomalous self-energy is given by
\begin{eqnarray}
\Sigma_{ll'}^a(\e_n)&=& \frac{n_{\rm imp}}{N}
\sum_{\k,m_i}T_{lm_1}(\e_n)G_{m_1m_2}(\k,\e_n)
\Delta_{m_2m_3}(\k,\e_n)
\nonumber \\
& &\times G_{m_4m_3}(-\k,-\e_n)T_{l'm_4}(-\e_n).
\label{eqn:SigmaA}
\end{eqnarray}
Then, the Eliashberg equation for $n_{\rm imp}\ne0$ is given by
using the full Green function 
${\hat G}(k)=[i\e_n+\mu-{\hat H}_\k^0-{\hat \Sigma}^n(\e_n)]^{-1}$
in eqs. (\ref{eqn:Eliash}) and (\ref{eqn:SigmaA}),
and adding $\Sigma_{ll'}^a(\e_n)$ to the right hand side of 
eq. (\ref{eqn:Eliash}).
Hereafter, we solve the equation at relatively high temperature $T=0.02$ 
since the number of $\k$-meshes ($128^2$) is not enough for $T<0.02$, 
due to the fact that $k_{\rm F}$ in iron pnictides is only 1/5 of that 
in cuprate superconductors.

\begin{figure}[!htb]
\includegraphics[width=0.9\linewidth]{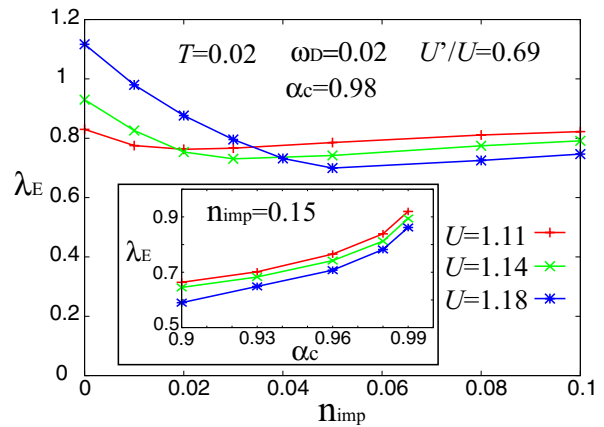}
\caption{
(Color online) 
$n_{\rm imp}$-dependence of $\lambda_{\rm E}$ at $\a_{\rm c}=0.98$.
If we put $g(0)=0$ ($s_\pm$-state),
$\lambda_{\rm E}$ at $n_{\rm imp}=0$ decreases by $0.1\sim0.15$,
since the ferro-orbital fluctuations enhance both 
$s_{++}$ and $s_\pm$ wave states.
Inset: $\a_{\rm c}$-dependence of $\lambda_{\rm E}$.
}
\label{fig:lambda}
\end{figure}

Figure \ref{fig:lambda} shows the $n_{\rm imp}$-dependence of 
$\lambda_{\rm E}$ at $\a_{\rm c}=0.98$, 
for $U=1.11$, 1.14 and 1.18.
Considering large $\lambda_{\rm E}\gtrsim0.8$ at $T=0.02$, 
relatively high-$T_{\rm c}$ ($\lesssim0.02$) is expected.
For the smallest $U$ ($U=1.11$; $\a_{\rm s}=0.85$),
we find that nearly isotropic $s_{++}$-wave state 
is realized;
the obtained $\lambda_{\rm E}$ is almost independent of $n_{\rm imp}$,
indicating the absence of impurity effect on the $s_{++}$-wave state,
as discussed in Refs. \cite{Onari-impurity,comment2}.
For the largest $U$ ($U=1.18$; $\a_{\rm s}=0.91$),
$s_\pm$-wave state is realized at $n_{\rm imp}=0$;
$\lambda_{\rm E}$ decreases slowly as $n_{\rm imp}$ 
increases from zero, whereas it saturates for $n_{\rm imp}\ge0.05$,
indicating the smooth crossover from $s_\pm$- to $s_{++}$-wave states
due to the interband impurity scattering.
For $U=1.14$ ($\a_{\rm s}=0.88$), the SC gap at $n_{\rm imp}=0$ is a hybrid of 
$s_{++}$ and $s_\pm$; only $\Delta_{\rm FS2}$ is different in sign.

The inset of Fig. \ref{fig:lambda} shows $\lambda_{\rm E}$ for 
$s_{++}$-wave state in the presence of impurities ($n_{\rm imp}=0.15$):
Since $\lambda_{\rm E}(\a_{\rm c}=0.98)-\lambda_{\rm E}(\a_{\rm c}=0.90)$
is only $\sim0.15$ for each value of $U$,
we expect that relatively large $T_{\rm c}$ for $s_{++}$-wave state
is realized even if orbital fluctuations are moderate.
We stress that the obtained $\lambda_{\rm E}$ is almost constant
for $\w_{\rm D}=0.02\sim0.1$,
suggesting the absence of isotope effect in the $s_{++}$-wave state 
due to the strong retardation effect \cite{Yada}.
By the same reason, $\lambda_{\rm E}$ for the the $s_{++}$-wave state
is seldom changed if we put $U=3$ in the Hartree-Fock term 
$\frac12({\hat \Gamma}^s-{\hat \Gamma}^c)$ in $W(q)$,
indicating that the Morel-Anderson pseudo-potential almost saturates.

\begin{figure}[!htb]
\includegraphics[width=0.9\linewidth]{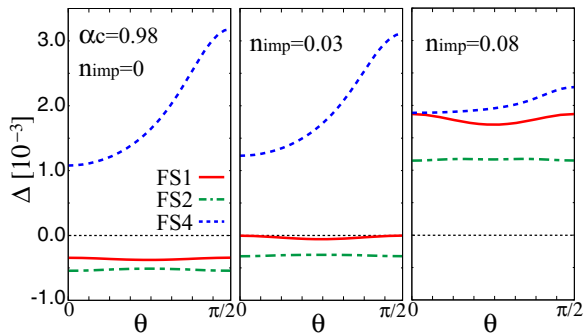}
\caption{
(Color online) 
SC gap functions for $U=1.18$ as functions of $\theta$
at (a) $n_{\rm imp}=0$, (b) 0.03, and (c) 0.08, respectively.
}
\label{fig:SCgap}
\end{figure}

Here, we discuss the case $U=1.18$ in detail:
Figure \ref{fig:SCgap} shows the SC gap 
on the FSs in the band-representation 
for (a) $n_{\rm imp}=0$, (b) 0.03, and (c) 0.08.
They satisfy the condition $N^{-1}\sum_{\k,lm}|\Delta_{lm}(\k)|^2=1$.
The horizontal axis is the azimuth angle for the $\k$-point
with the origin at $\Gamma$ (M) point for FS1,2 (FS4);
$\theta=0$ corresponds to the $k_x$-direction.
In case (a), $s_\pm$-state with strong imbalance,
$|\Delta_{\rm FS1}|, |\Delta_{\rm FS2}| \ll \Delta_{\rm FS4}$, is realized,
and $\Delta_{\rm FS4}$ takes the largest value at $\theta=\pi/2$,
where the FS is mainly composed of orbital 4.
In case (c), impurity-induced isotropic $s_{++}$-state \cite{Hirschfeld}
with $\Delta_{\rm FS1}\sim \Delta_{\rm FS2}\sim \Delta_{\rm FS34}$
is realized, consistently with many ARPES measurements \cite{ARPES}.
In case (b), $\Delta_\k$ on FS1 is almost gapless.
However, considering the $k_z$-dependence of the FSs, 
a (horizontal-type) nodal structure is expected to appear on FS1,2.
In real compounds with $T_{\rm c}\sim50$K,
the $s_\pm \rightarrow s_{++}$ crossover should be induced
by small residual resistivity $\rho_{\rm imp}\sim20\ \mu\Omega{\rm cm}$
($n_{\rm imp}\sim 0.01$ for $I=1$),
as estimated in Ref. \cite{Onari-impurity}.

We comment that at $n_{\rm imp}=0$, $s_\pm$-wave state is realized in the RPA 
even if $\a_{\rm s}\lesssim \a_{\rm c}$, due to factor 3 in front of 
$\frac12{\hat \Gamma}^s{\hat \chi}^s(q){\hat \Gamma}^s$ in $W(q)$.
For the same reason, however, reduction in $\a_{\rm s}$
(or increment of $U_{\rm cr}$ for $\a_{\rm s}=1$) due to the 
``self-energy correction by $U$'' is larger, 
which will be unfavorable for the $s_\pm$-wave state.
Therefore, self-consistent calculation for the self-energy is required 
to discuss the value of $\a_{\rm c,s}$ and the true pairing state.

Here, we discuss where in the $\a_{\rm s}$-$\a_{\rm c}$ phase diagram 
in Fig. \ref{fig:phase} (a) real compounds are located.
Considering the weak $T$-dependence of $1/T_1T$ 
in electron-doped SC compounds \cite{Hosono},
we expect that they belong to the area $\a_{\rm c}\gg\a_{\rm s}$.
Then, $s_{++}$-wave SC state will be realized
without (or very low density) impurities, like the case of
$U=1.11$ or 1.14 in Fig. \ref{fig:lambda}.
On the other hand, 
impurity-induced $s_\pm \rightarrow s_{++}$ crossover may be realized
in BaFe$_2$(As$_{1-x}$P$_x$)$_2$ (undoped) or 
(Ba$_{1-x}$K$_{x}$)Fe$_2$As$_2$ (hole-doped) SC compounds,
where AFM fluctuations are rather strong.

Finally, we discuss the non-Fermi-liquid-like transport phenomena 
in iron pnictides.
For example, the resistivity is nearly linear-in-$T$, and 
the Hall coefficient $R_{\rm H}$ increases at lower temperatures 
\cite{Sato-imp,RH}.
Although the forward scattering induced by ferro-orbital fluctuations 
might be irrelevant,
antiferro-orbital and AFM fluctuations with $\Q=(\pi,0)$
are expected to cause the anomalous transport,
due to the current vertex correction \cite{ROP}.

In summary, we have proposed a mechanism of $s_{++}$-wave SC state
induced by orbital fluctuations, due to the
phonon-mediated electron-electron interaction.
Three orbitals ($XZ$, $YZ$ and $X^2-Y^2$) are 
necessary to leads the ferro-orbital fluctuations.
The SC gap structure drastically changes depending on parameters
$\a_{\rm s}$, $\a_{\rm c}$, and $n_{\rm imp}$, 
consistently with observed rich variety of the gap structure
that is a salient feature of iron pnictides.
Orbital fluctuation mediated $s_{++}$-wave state 
is also obtained for hole-doped cases,
although the antiferro-orbital fluctuations
becomes stronger than the ferro-orbital ones.

The $s$-wave superconductivity induced by orbital fluctuations
had been discussed in Ref. \cite{Takimoto} for $U'>U$;
this condition can be realized by including the $A_{1g}$-phonon \cite{Yanagi2}.
In the present model, however, $A_{1g}$-phonon is negligible since 
$g_{\rm cr}(0)$ given by $A_{1g}$-phonon 
is much greater than $g_{\rm cr}(0)\sim0.4$ in Fig. \ref{fig:phase} (a):
The ferro-obtital fluctuations in Fig. \ref{fig:phase} (b) originate from
the {\it negative} exchange interaction caused by $E_{g}$-phonon,
as shown in Fig. \ref{fig:V} (c).

\acknowledgements
We thank D.S. Hirashima, M. Sato, Y. Matsuda, Y. {\=O}no
and Y. Yanagi for valuable discussions.
This study has been supported by Grants-in-Aid for Scientific 
Research from MEXT of Japan, and by JST, TRIP.

Note added in proof:
After the acceptance of this work,
we found that $g_{\rm cr}(0)\sim0.4$ in Fig. \ref{fig:phase} (a)
is reduced to half if all the $e$-ph matrix elements 
including $1, 5$ orbitals are taken into account.
Results similar to Fig. \ref{fig:lambda} are obtained 
by using $g(0)\sim0.2$, whereas (vertical-type) nodes appear on FS3,4 
during the $s_{++}\rightarrow s_\pm$ crossover for $U=1.18$.



\begin{thebibliography}{99}

\bibitem{Kuroki}
K. Kuroki {\it et al}., 
Phys. Rev. Lett. {\bf 101}, 087004 (2008).

\bibitem{Mazin}
I. I. Mazin {\it et al}., 
Phys. Rev. Lett. {\bf 101}, 057003 (2008).

\bibitem{Onari-impurity}
S. Onari and H. Kontani, 
Phys. Rev. Lett. {\bf 103} 177001 (2009).

\bibitem{Sato-imp}
A. Kawabata {\it et al.}, J. Phys. Soc. Jpn. {\bf 77} (2008) Suppl. C 103704;
M. Sato {\it et al.}, J. Phys. Soc. Jpn. {\bf 79} (2009) 014710;
S.C. Lee {\it et al.}, J. Phys. Soc. Jpn. {\bf 79} (2010) 023702.

\bibitem{irradiation}
C. Tarantini {\it et al.}, arXiv:0910.5198.

\bibitem{christianson}
A. D. Christianson, {\it et al}., Nature {\bf 456}, 930 (2008).

\bibitem{Onari-resonance}
S. Onari {\it et al}, Phys. Rev. B {\bf 81}, 060504(R) (2010).

\bibitem{lambda-LDA}
L. Boeri {\it et al.}, Phys. Rev. Lett. {\bf 101}, 026403 (2008).

\bibitem{softening}
 R.M. Fernandes {\it et al.}, arXiv:0911.3084.

\bibitem{Raman}
M. Rahlenbeck {\it et al.}, Phys. Rev. B {\bf 80}, 064509 (2009).




\bibitem{AsP}
K. Hashimoto {\it et al.}, arXiv:0907.4399.

\bibitem{Izawa}
C. Martin et al, Phys. Rev. B {\bf 81}, 060505(R) (2010).

\bibitem{Scalapino}
T.A. Maier, {\it et al.}, Phys. Rev. B {\bf 79}, 224510 (2009). 

\bibitem{Yada}
K. Yada and H. Kontani, Phys. Rev. B {\bf 77}, 184521 (2008).


\bibitem{Singh}
D.J. Singh, Physica C {\bf 469}, 418 (2009). 

\bibitem{gunnarsson}
J.E. Han {\it et al.}, Phys. Rev. Lett. {\bf 90}, 167006 (2003);
M. Capone {\it et al.}, Phys. Rev. Lett. {\bf 93}, 047001 (2004).

\bibitem{Takimoto}
 T. Takimoto {\it et al.}, 
 J. Phys.: Condens. Matter {\bf 14}, L369 (2002). 

\bibitem{comment}
The effect of Coulomb interaction on $\chi^c_{24,42}(\q,0)$ 
is not large if $C_{ll',ll'}+C_{ll',l'l}=-U'+J+J'$ is small.

\bibitem{comment3}
$\lambda_i$ for orbital $i=2\sim4$ is 
$\lambda_i\approx-\sum_{j=2}^4N_j(0)V_{ij,ij}(0)=N(0)g(0)$,
where $N_j(0)$ is the partial DOS.
Then, $\lambda\approx N(0)g(0)$ in the band-diagonal basis.


\bibitem{Shimo-OO}
T. Shimojima {\it et al}, Phys. Rev. Lett. {\bf 104}, 057002 (2010).

\bibitem{OO-theory}
 F. Kr{\"u}ger {\it et al.}, Phys. Rev. B {\bf 79}, 054504 (2009);
 W. Lv {\it et al.}, Phys. Rev. B {\bf 80}, 224506 (2009);
 C.C. Lee {\it et al.}, Phys. Rev. Lett. {\bf 103}, 267001 (2009).

\bibitem{comment2}
Above $T_{\rm c}$,
$\lambda_{\rm E}$ slightly {\it increases with $n_{\rm imp}$}
in conventional $s$-wave superconductors, 
but never exceeds unity.

\bibitem{Hirschfeld}
V. Mishra {\it et al.}, Phys. Rev. B {\bf 79}, 094512 (2009);
D. Markowitz {\it et al.}, Phys. Rev. {\bf 131}, 563 (1963).

\bibitem{ARPES}
D. V. Evtushinsky {\it et al}., New J. Phys. {\bf 11}, 055069 (2009).


\bibitem{Hosono}
T. Nakano {\it et al.}, Phys. Rev. B {\bf 81}, 100510(R) (2010);
Y. Nakai {\it et al.}, Phys. Rev. B {\bf 81}, 020503(R) (2010).


\bibitem{RH}
S. Kasahara {\it et al.}, arXiv:0905.4427.

\bibitem{ROP}
H. Kontani, Rep. Prog. Phys. {\bf 71}, 026501 (2008).



\bibitem{Yanagi2}
Y. Yanagi {\it et al.}, 
Phys. Rev. B {\bf 81}, 054518 (2010).


\end{thebibliography}
\end{document}